\documentstyle[12pt,epsf]{article}
\newlength{\abstractwidth}
\renewcommand{\thefootnote}{\fnsymbol{footnote}}
\renewcommand{\thanks}[1]{\footnote{#1}} % Use this for footnotes
\newcommand{\starttext}{
\setcounter{footnote}{0}
\renewcommand{\thefootnote}{\arabic{footnote}}}
\renewcommand{\theequation}{\thesection.\arabic{equation}}
\newcommand{\be}{\begin{equation}}
\newcommand{\bea}{\begin{eqnarray}}
\newcommand{\eea}{\end{eqnarray}}
\newcommand{\beq}{\begin{equation}}
\newcommand{\ee}{\end{equation}}
\newcommand{\eeq}{\end{equation}}
\newcommand{\N}{{\cal N}}

\def\ba{\begin{eqnarray}}
\def\ea{\end{eqnarray}}

\def\N{{\cal N}}

\def\12{{1 \over 2}}
\def\32{{3 \over 2}}
\def\72{{7 \over 2}}
\def\92{{9 \over 2}}

\def\nc{non--commutative}

\def\ft{fuzzy torus}
\def\udag{U^{\dag}}
\def\vdag{V^{\dag}}
\def\xdag{X^{\dag}}
\def\ydag{Y^{\dag}}
\def\zdag{Z^{\dag}}
\def\ep{e^{i \theta}}
\def\em{e^{-i \theta}}
\def\epp{e^{2i \theta}}
\def\epn{e^{i(N-1) \theta}}
\def\q{&=&}
\def\x{{\cal{X}}}
\def\y{{\cal{Y}}}
\def\xd{{\cal{X}}^{\dag}}
\def\yd{{\cal{Y}}^{\dag}}
\def\wl{Wilson line}
\def\da{^{\dag}}
\begin{document}
\renewcommand{\theequation}{\thesection.\arabic{equation}}
\begin{titlepage}
\bigskip
\rightline{}  \rightline{hep-th/0109018}

\bigskip\bigskip\bigskip\bigskip

\centerline{\Large \bf {Gauge Theory On The Fuzzy Torus }}

\bigskip\bigskip
\bigskip\bigskip
%\centerline{\it }
%\medskip
%\centerline{} \centerline{} \centerline{}
%\bigskip

\centerline{\it Daniela Bigatti  }
\medskip
\centerline{Weizmann Institute of Science}
\centerline{P.O.Box 26 } 
\centerline{Rehovot 76100, Israel}
\bigskip\bigskip
\begin{abstract}
In this paper a formulation of U(1) gauge theory on a fuzzy torus is
discussed.  The theory is regulated in both the infrared and
ultraviolet. It can be thought of as a non-commutative version of
lattice gauge theory on a periodic lattice. The construction of Wilson
Loops is
particularly transparent in this formulation. Following
Ishibashi,  Iso,  Kawai and Kitazawa,
we show that certain Fourier modes of open Wilson lines are
Gauge invariant.

We also introduce charged matter fields which can be thought of as
fundamentals of the gauge group. These particles  behave like
charges in a strong magnetic field and are frozen into the lowest
Landau levels. The resulting system is a simple matrix quantum
mechanics which should reflect much of the physics of charged particles
in
strong magnetic fields. 

The present results were first presented as a talk at the Institute for Mathematical Science, Chennai, India; the author wishes to thank Prof.~T.~R.~Govindarajan and the IMS for hospitality and financial support, and the audience for pointing me out the work of 
Ambjorn,  Makeenko,  Nishimura and Szabo. The author is also grateful to the organizers of the Summer School in Modern Mathematical Physics, Sokobanja, Serbia, Yugoslavia, for financial support and hospitality during presentation of the present work at international conference FILOMAT 2001, Ni\v s, Yugoslavia.

\medskip
\noindent
\end{abstract}
\end{titlepage}
\starttext \baselineskip=18pt \setcounter{footnote}{0}

%%%%%%%%%%%%%%%%%%%%%%%%%%%%%%%%%%%%%%%%%%%%%%%%%%%%%%%%%%%%%%%%%%%%%%
%%%%%%
%%%%%%%%%%%%%%%%
%%%%%%%%%%%%%%%%%%%%%%%%%%%%%%%%%%%%%%%%%%%%%%%%%%%%%%%%%%%%%%%%%%%%%%
%%%%%%
%%%%%%%%%%%%%%%%
\setcounter{equation}{0}
\section{The Fuzzy Torus}

In this paper we will be interested in the construction of gauge
invariant Wilson Loops in a regularized version of \nc \ gauge
theory. After  this paper was written we realized that the theory we are using and much of our results  have previously been discussed by
Ambjorn,  Makeenko,  Nishimura and
Szabo \cite{1}. Our presentation  is much less general than theirs but
because it is also very simple we felt it was worth circulating.

The regularized theory is a \nc \ version of lattice gauge
theory on the Fuzzy Torus. It is patterned after the the
Hamiltonian form of lattice gauge theory  \cite{2}.

The lattice version is an especially
intuitive formulation of the non-perturbative theory.
For illustrative purposes we will concentrate on the
Abelian theory in $2+1$ dimensions. The generalization to higher
dimensions and non-abelian gauge groups is straight forward.
Our main focus will be on defining the gauge
invariant quantities of the theory including closed and open
Wilson lines and in formulating the theory of matter in the
fundamental representation of the noncommutative algebra of
functions.

The fuzzy 2-torus is defined by non-commuting coordinates $U,V$
satisfying
\bea
U^{\dag}U&=&V^{\dag}V =1 \cr
U^N&=&V^N =1 \cr
UV &=& VU e^{i \theta}
\eea
with $\theta = {2 \pi} /N $. These relations can be represented
by $N \times N$ matrices
\bea
U =\left (\begin{array}{cccccccc}0&1&0&0&0&0&.&. \\0&0&1&0&0&0&.&.
\\0&0&0&1&0&0&.&. \\.&.&.&.&.&.&.&. \\1&0&0&0&0&0&.&.
\end{array}\right)
\eea
\bea
V =\left (\begin{array}{cccccccc}1&0&0&0&0&0&.&.
\\0&{e^{i\theta}}&0&0&0&0&.&.
\\0&0&{e^{2i\theta}}&0&0&0&.&.
\\.&.&.&.&.&.&.&. \\0&0&0&0&0&0&.&{e^{i(N-1)\theta}}
\end{array}\right)
\eea

The physical interpretation of $U,V $ is that they are
exponentials   of non-commuting periodic coordinates
\bea
U&=&\exp{ i x \over R} \cr
V&=&\exp{ i y \over R}
\eea
Formal manipulations would indicate nontrivial commutation
relations for $x,y$ of the form
\be
[y,x]=i\theta R^2
\ee
This relation can be satisfied by introducing a pair of operators
on Hilbert space $q,p$ with
\be
p=-i\partial_q
\ee
and defining
\bea
y \q qR {\theta}^{1\over 2} \cr
x \q pR {\theta}^{1\over 2}
\eea

This equation is sometimes a useful mnemonic but it is not
strictly correct;  no two finite dimensional matrices can have a
commutator with a non-vanishing trace.

Functions on the \ft \ are defined by the \nc analogue of a Fourier
series.
\be
\phi(U,V)=\sum_{n,m=0}^{N-1} c_{mn}U^nV^m
\ee
It is convenient to define
\bea
U_{mn}&=&e^{-imn\theta} U^mV^n \cr
\phi_{mn} &=&e^{imn\theta}c_{mn}\cr
\phi(U,V)&=&\sum_{n,m=0}^{N-1}\phi_{mn}U_{mn}
\eea
Note that the $U_{mn} $ satisfy
\be
U_{mn}U_{rs}=\exp{\12 i{\theta }(ms-nr)} \equiv U_{mn}*U_{rs}
\ee
Equation (1.8) defines the star-product on the \ft .

The \ft  \ is analogous to a periodic lattice with a spacing

\be
 a=2 \pi
R/N.
\ee
 This is because the Fourier expansion
in eq.(1.7) has only a finite number of terms. In other words
there is a largest momentum in each direction
\be
p_{max} = 2\pi(N-1)/R
\ee
Thus the \ft \ has both an infrared cutoff length $R$ and an
ultraviolet cutoff length $2 \pi
R/N$

The operators $U,V$ function as shifts  on the periodic lattice.
Using the last of eq's(1.1) one easily finds
\bea
U\phi(U,V)\udag&=&\phi(U, Ve^{i \theta})\cr
\udag \phi(U,V)U&=&\phi(U, Ve^{-i \theta})\cr
V\phi(U,V)\vdag&=&\phi(Ue^{-i \theta}, V)\cr
\vdag \phi(U,V)V&=&\phi(Ue^{i \theta}, V)\cr
\eea
More generally
\be
U^n V^m \phi(U,V)
{\vdag}^m {\udag}^n =
\phi(U e^{-im\theta}, V e^{in\theta} )
\ee

The rule for integration on the \ft \ is simple.
\be
\int U_{mn} = 4 {\pi}^2 R^2 \delta_{m0} \delta_{n0}
\ee
Noting that
\be
Tr U_{mn} =N\delta_{m0} \delta_{n0}
\ee
we make the identification
\be
\int F(U,V) = {4 {\pi}^2 R^2 \over N } Tr F(U,V)
\ee
%%%%%%%%%%%%%%%%%%%%%%%%%%%%%%%%%%%%%%%%%%%%%%%%%%%%%%%%%%%%%%%%%%%%%%%%%%%

%%%%%%%%%%%%%%%%%%%%%%%%%%%%%%%%%%%%%%%%%%%%%%%%%%%%%%%%%%%%%%%%%%%%%%%%%%%

%%%%%%%%%%%%%%%%%%%%%%%%%%%%%%%%%%%%%%%%%%%%%%%%%%%%%%%%%%%%%%%%%%%%%%%%%%%

\setcounter{equation}{0}
\section{Gauge Theory  On The Fuzzy Torus}

In what follows  we will work in the temporal gauge in which the
time component of the vector potential is zero.

Let us introduce gauge fields on the fuzzy torus in analogy with
the link variables of lattice gauge theory \cite{2}. We will explicitly
work with the gauge group $U(1)$. The link variable in the $x,y$
direction is called $X,Y$. The link variables are unitary
\bea
\xdag X&=&1 \cr
\ydag Y&=&1
\eea
The gauge invariance of the theory is patterned on that of lattice
gauge theory. Let $Z$ be a unitary, time independent  function of $U,V$,

$\zdag
Z=1$. The gauge transformation induced by $Z$ is defined to be
\bea
X'&=&Z(U,V)X(U,V) \zdag (U\ep , V) \cr
Y'&=&Z(U,V)Y(U,V) \zdag (U , V\ep)
\eea
or
\bea
X'&=&Z X \ \vdag \zdag V \cr
Y'&=&ZY U \zdag \udag
\eea

Let us now construct Wilson loops by analogy with the conventional
lattice construction. We will give some examples first. A Wilson
line which winds around the x-cycle of the torus at a fixed value
of $y$ is given by
\bea
W_x &=& Tr X(U,V)X(U\ep,V)X(U\epp,V)..X(U\epn,V)\cr
    &=& Tr (X\vdag)^N
\eea
Similarly
\be
W_y=Tr (YU)^N
\ee
These expressions are gauge invariant under the transformation in
eq.(2.3).

Another example of a Wilson loop is the analogue of the plaquette
in lattice gauge theory. It is given by
\bea
{\cal{P}} \q Tr X(U,V)Y(U\ep ,V)\xdag(U,V\ep)\ydag(U,V)\cr
\q Tr (X)(\vdag Y V)(U\xdag \udag)( \ydag)\cr
\q \em Tr(X\vdag)(YU)(V\xdag)(\udag \ydag)
\eea

The general rule involves drawing a closed  oriented chain formed
from directed links. A step in the positive (negative) $x$ direction is
described by the link operator $X\vdag$ ($V \xdag$). Similarly a step in

the
positive (negative) $y$ direction gives a factor $YU$ ($\udag
\ydag$). The link operators are multiplied in the order specified
by the chain and the trace is taken. In addition there is a factor
$e^{-i A
\theta}$ where $A$ is the signed Area of the loop in units of the
lattice spacing. For a simple  contractable clockwise oriented loop with

no
crossings, $A$ is just the number of enclosed plaquettes.

A simple Lagrangian for the gauge theory  can be formed from
plaquette operators and kinetic involving time derivatives. The
expression
\be
Tr \dot {\xdag} \dot {X}+\dot {\ydag} \dot {Y}
\ee
is quadratic in time derivatives and is gauge invariant.
Again, following the model of lattice gauge theory [KS] we choose the
action
\be
{\cal{L}} =
{4\pi^2 R^2 \over g^2 a^2 N}
Tr\left [ \dot {\xdag} \dot {X}+\dot {\ydag} \dot {Y}
+{\em \over a^2}  (X\vdag)(YU)(V\xdag)(\udag \ydag)+cc \right]
\ee

Evidently the operators $X \vdag$ and $YU$ play an important
role.  We therefore define
\bea
\x \q X\vdag \cr
\y \q YU
\eea
These operators transform simply under gauge transformations:
\bea
\x &\to& Z\x Z^{\dag} \cr
\y  &\to& Z\y Z^{\dag}
\eea

The action is now written in the form
\be
{\cal{L}}={4\pi^2 R^2 \over g^2 a^2 N}
Tr\left [ \dot {\xd} \dot {\x}+\dot {\yd} \dot {\y}
+{\em \over a^2} {\x}{\y}{\xd} {\yd}+cc \right]
\ee
or using eq.(1.9)
\be
{\cal{L}}={N \over g^2}
Tr\left [ \dot {\xd} \dot {\x}+\dot {\yd} \dot {\y}
+{\em \over a^2} {\x}{\y}{\xd} {\yd}+cc \right]
\ee
In this form the action is equivalent to that of a $U(N)$ lattice
gauge theory formulated on a single plaquette but with periodic
boundary conditions of a torus. This appears to be a form of
Morita equivalence \cite{3}.

If the coupling constant is small, the ground state is determined
by minimizing the plaquette term in the Hamiltonian. This is done
by setting
\be
\em {\x}{\y}{\xd} {\yd}=1
\ee
Up to a gauge transformation the unique solution of this equation
is
\bea
\x &=&\vdag  \cr
\y \q U
\eea
or
\be
X=Y=1
\ee

\setcounter{equation}{0}
\section{Open Wilson Loops}

Thus far we have constructed closed Wilson loops. Recall that  the
construction involves taking a trace. This is the analogue of
integrating the location of the Wilson loop over all space. In
other words the closed Wilson Loop carries no spatial momentum. In
a very interesting paper Ishibashi,  Iso,  Kawai and Kitazawa
\cite{4} have argued that there
exist gauge invariant operators which correspond to specific
Fourier modes of open Wilson lines. These objects are very closely
related to the growing dipoles of \nc \ field theory whose size
depends on their momentum \cite{5,6}. Das and Rey \cite{7} have shown
that these
operators are a complete set of gauge invariant operators. Their
importance has been further clarified by  Gross,  Hashimoto and Itzhaki
\cite{8}.

Let us consider the simplest example of an open Wilson line, ie, a
single  link variable, say $X$. From eq(2.3) we see that $X$ is
not gauge invariant.  But now consider $X \vdag =\x$. Under gauge
transformations
\be
\x  \to Z\x Z^{\dag}
\ee
Evidently the quantity
\be
Tr X \vdag =Tr \x
\ee
is gauge invariant. Now using eq(1.4) we identify this quantity as
\be
Tr X \vdag ={N\over 4 \pi^2 R^2}\int X e^{-iy\over R} d^2 x
\ee
Thus we see that a particular Fourier mode of $X$ is gauge
invariant.

Let us consider another example in which an open Wilson line
consist of two adjacent links, one along the $x$ axis and one
along the $y$ axis.
\be
X \vdag YV =\x \y \udag V
\ee
Multiplying by $\vdag U$ and taking the trace gives
\be
Tr(X \vdag YV)\vdag U = Tr \x \y = gauge \ invariant
\ee
But we can also write this as
\be
{N\over 4 \pi^2 R^2}\int d^2x (X \vdag YV)e^{-iy\over R}e^{-ix\over R}
\ee
In other words it is again a Fourier mode of the open Wilson line.
In general the particular Fourier mode is related to the
separation between the endpoints of the \wl by the same relation
as that in \cite{5,6} where it was shown that a particle in \nc \ field
theory is a dipole oriented perpendicular to it momentum with a
size proportional to the momentum.

\setcounter{equation}{0}
\section{Fields in the Fundamental Representation}

In this section we will define fields in the fundamental
representation of the gauge group. For simplicity we consider
non-relativistic
particles. Let us begin with what we do
$not$ mean particles in the fundamental. Define a complex valued
field $\phi$ that takes values in the $N \times N$ dimensional matrix
algebra
generated by $U,V$. The gauge transformation properties of $\phi$ are
given by
\be
\phi  \to Z \phi.
\ee
Note that this is left multiplication by $Z$ and not conjugation.
The field $\phi$ carries a single unit of abelian gauge charge. Although
the field
has two indices in the $N$ dimensional space the gauge transformations
only act on the
left index.

An obvious choice of gauge invariant
"hopping"
Hamiltonian would be
\be
H \sim   Tr \phi \da X \vdag \phi V  \phi \da X U \phi \udag +cc
\ee
In a non-abelian theory a similar construction can be carried out
for quark fields in the fundamental.

We shall mean something different  by fields in the fundamental.
Such fields have only $one$ index. They are vectors rather than matrices

in the Hilbert space that the represents
the algebra of functions. In the present case they are $N$
component complex vectors $|\psi\rangle$. These fields represent
particles moving in a strong magnetic field which are frozen into
the lowest Landau level.

Consider the case of non-relativistic particles moving on
the \nc \ lattice. The conventional lattice action would be
\be
L=L_0 - L_h
\ee
Where
\be
L_0 = i\left(\langle\dot \psi \da |\psi \rangle  -cc     \right)
\ee
and  $L_h$ is a hopping Hamiltonian. The natural \nc \ version of
the hopping term is
\be
L_h ={1 \over a} \langle \psi| X \vdag +YU -2 |\psi\rangle + cc
\ee
The presence of the link variables $X,Y$ is familiar from ordinary
lattice field theory and the $\vdag , U$ are  the shifts which
move $\psi$. We may also write the hopping term as
\be
L_h = {1\over a} \langle \psi| \x +\y | -2 \psi\rangle + cc
\ee

Combining (2.12), (4.2) and (4.4)

\be
{\cal{L}}={N \over g^2}
Tr\left [ \dot {\xd} \dot {\x}+\dot {\yd} \dot {\y}
+{\em \over a^2} {\x}{\y}{\xd} {\yd}+cc \right]
+ i\langle\dot \psi \da |\psi \rangle
+ {1\over a }\langle \psi| \x +\y -2 |\psi\rangle + cc
\ee
Let us consider hopping terms in (4.6). In the limit of weak coupling
we may use eq.(2.14)
to give
\be
L_h= {1 \over a}\langle \psi| \vdag +U -2 |\psi\rangle + cc
\ee
To get some idea of the meaning of this term let us use eq(1.4)
and expand the exponentials.
\be
L_h= {1 \over a} \langle \psi| {(x^2 +y^2)\over R^2 }|\psi\rangle + cc
\ee
and using (1.7)
\be
L_h= {1 \over a} \langle \psi| (p^2+q^2 )\theta|\psi\rangle + cc
\ee
Thus we recognize this term as a
harmonic oscillator hamiltonian with an in spectrum of levels spaced by
$\theta \sim N^{-1}$.
 Evidently, in this approximation
the particles move
in quantized circular orbits around the origin.

This phenomena is related to the fact that the fundamental
particles behave like charged particles in a strong magnetic field
and are frozen into their lowest Landau levels. Furthermore the
LLL's are split by a force attracting the particles to $x=y=0$.
This has a natural interpretation in matrix theory in which the
same system appears as a 2-brane and 0-brane with strings
connecting them \cite{9}.

\setcounter{equation}{0}
\section{Rational Theta}

Thus far we have worked with eq.(1.1) with $\theta =2 \pi /N$. Let
us generalize the construction to the case $\theta = 2 \pi p/N$
with $p$ relatively prime to $N$. We continue to define the fuzzy
torus by eq.(1.1). Let us define two  matrices $u,v$ satisfying
\bea
u^{\dag}u&=&v^{\dag}v =1 \cr
u^N&=&v^N =1 \cr
uv &=& vu e^{2 \pi i \alpha \over N}
\eea
such that
\be
\alpha p =1(mod\N)
\ee
Then it follows that
\bea
U \q u^p \cr
V \q v^p
\eea
satisfies eq.(1.1). Furthermore, $u$ and $v^{\dag}$ act as shifts
by distance $2 \pi R/N$;
\bea
u V u^{\dag} \q V \exp\left ({2 \pi i \over N}\right) \cr
v^{\dag} U v \q U \exp\left ({2 \pi i \over N}\right)
\eea
The basic plaquette is now given by
\bea
{\cal{P}} \q Tr(X)(v^{\dag}Y v)(u \xdag u^{\dag})(\ydag) \cr
\q  e^{2 \pi i \alpha \over N} Tr(X v^{\dag})(Yu) (v\xdag)(u^{\dag}
\ydag)
\eea

The final expression for action is essentially the same as in
eq.(2.10) except that the factor $\ep$ is replaced by
$ e^{2 \pi i \alpha \over N}$.

We can now describe one approach to the continuum limit, $a \to 0$. To
get to
such a limit  eq.(1.9) requires that $N \to \infty$. We also
want the theta parameter to approach a finite limit. This requires
$p/N$ to approach a limit. for example if we want $p/N \to 1/2$ we
can choose the sequence
($p =n, N=2n+1$) so that $p$ and $N$ remain relatively prime. In
this way $p/N$ can tend to a rational or irrational limit and the
lattice spacing will approach zero.
%%%%%%%%%%%%%%%%%%%%%%%%%%%%%%%%%%%%%%%%%%%%%%%%%%%%%%%%%%%%%%%%%%%%%%%%%%%%

%%%%%
\section*{Acknowledgements}

The present results were first presented as a talk at the Institute for Mathematical Science, Chennai, India; the author wishes to thank Prof.~T.~R.~Govindarajan and the IMS for hospitality and financial support, and the audience for pointing me out the work of 
Ambjorn,  Makeenko,  Nishimura and Szabo. The author is also grateful to the organizers of the Summer School in Modern Mathematical Physics, Sokobanja, Serbia, Yugoslavia, for financial support and hospitality during presentation of the present work at international conference FILOMAT 2001, Ni\v s, Yugoslavia.

%%%%%%%%%%%%%%%%%%%%%%%%%%%%%%%%%%%%%%%%%%%%%%%%%%%%%%%%%%%%%%%%%%%%%%%%%%%%

%%%%%
%%%%%%%%%%%%%%%%%%%%%%%%%%%%%%%%%%%%%%%%%%%%%%%%%%%%%%%%%%%%%%%%%%%%%%%%%%%%

%%%%%
\end{document}